\documentclass[conference]{IEEEtran}
\IEEEoverridecommandlockouts
\usepackage{cite}
\usepackage{float}
\usepackage{longtable}
\usepackage{supertabular}
\usepackage{tabularx}
\usepackage{amsmath,amssymb,amsfonts}
\usepackage{algorithmic}
\usepackage{graphicx}
\usepackage{textcomp}
\usepackage{xcolor}
\usepackage{makecell}
\usepackage{comment}
\usepackage{ragged2e}
\usepackage{blindtext}
\def\BibTeX{{\rm B\kern-.05em{\sc i\kern-.025em b}\kern-.08em
    T\kern-.1667em\lower.7ex\hbox{E}\kern-.125emX}}
\begin{document}

\title{Application of Explainable Machine Learning in Detecting and Classifying Ransomware Families Based on API Call Analysis\\

\thanks{This work is funded by NetApp.}
}

\author{\IEEEauthorblockN{Rawshan Ara Mowri}
\IEEEauthorblockA{\textit{Department of Computer Science} \\
\textit{North Carolina A\&T State University}\\
Greensboro, USA \\
rmowri@aggies.ncat.edu}
\and
\IEEEauthorblockN{Madhuri Siddula}
\IEEEauthorblockA{\textit{Department of Computer Science} \\
\textit{North Carolina A\&T State University}\\
Greensboro, USA \\
msiddula@ncat.edu}
\and
\IEEEauthorblockN{Kaushik Roy}
\IEEEauthorblockA{\textit{Department of Computer Science} \\
\textit{North Carolina A\&T State University}\\
Greensboro, USA \\
kroy@ncat.edu}
}

\maketitle

\begin{abstract}
Ransomware has appeared as one of the major global threats in recent days. The alarming increasing rate of ransomware attacks and new ransomware variants intrigue the researchers to constantly examine the distinguishing traits of ransomware and refine their detection strategies. Application Programming Interface (API) is a way for one program to collaborate with another; API calls are the medium by which they communicate. Ransomware uses this strategy to interact with the OS and makes a significantly higher number of calls in different sequences to ask for taking action. This research work utilizes the frequencies of different API calls to detect and classify ransomware families. First, a Web-Crawler is developed to automate collecting the Windows Portable Executable (PE) files of 15 different ransomware families. By extracting different frequencies of 68 API calls, we develop our dataset in the first phase of the two-phase feature engineering process. After selecting the most significant features in the second phase of the feature engineering process, we deploy six Supervised Machine Learning models: Naïve Bayes, Logistic Regression, Random Forest, Stochastic Gradient Descent, K-Nearest Neighbor, and Support Vector Machine. Then, the performances of all the classifiers are compared to select the best model. The results reveal that Logistic Regression can efficiently classify ransomware into their corresponding families securing 99.15\% overall accuracy. Finally, instead of relying on the ‘Black box’ characteristic of the Machine Learning models, we present the post-hoc analysis of our best-performing model using 'SHapley Additive exPlanations' or SHAP values to ascertain the transparency and trustworthiness of the model’s prediction.
\end{abstract}

\begin{IEEEkeywords}
Ransomware Classification, Machine Learning, Explainable AI, Cyber Security
\end{IEEEkeywords}

\section{Introduction}
Recently, ransomware has become one of the biggest global challenges that are agitating peoples’ normal lives. Being harmful software, it applies symmetric and asymmetric cryptography to inscribe user information and poses a Denial-of-Service (DoS) attack on the intended user [1]. The unique functional process of ransomware attacks makes it more harmful than any malware attacks and causes irreversible losses. According to [2], Fig. 1 illustrates the number of publicized ransomware attacks in 2021, with inflation of 25\% than the same time in the previous year. Although the report does not include the number of supply chain attacks, it is creating a big interference in providing healthcare, purchasing groceries, and even loading fuel in vehicles. Examples of these attacks are the Kaseya attack, the colonial pipeline attack, etc. In addition, in the first six months of 2021, the FBI’s Internet Crime Complaint Center documented 2,084 ransomware attacks [3], and the U.S. Treasury's Financial Crimes Enforcement Network (FinCEN) recorded the cost of around \$590 million related to ransomware activities during that period [4]. Moreover, distinct ransomware variants are being detected regularly and more than 130 different ransomware variants have been identified from 2020 till this year causing an inevitable disturbance in day-to-day lives. [5].

\begin{figure}[htb]
	\makebox[\linewidth][c]{\includegraphics[angle = 0, clip, trim=0cm 0cm 0cm 0cm, width=0.5\textwidth]{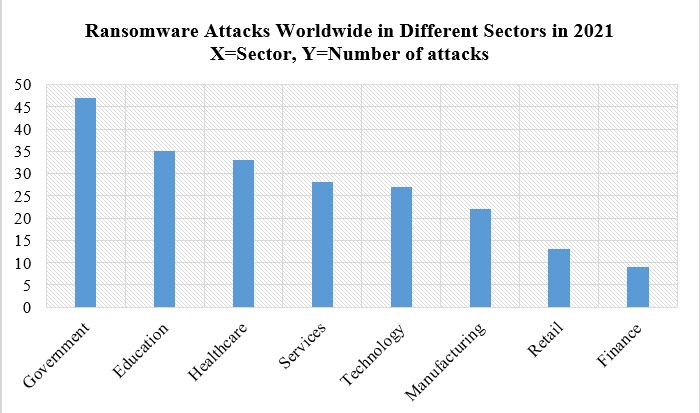}}
	\caption{Number of worldwide ransomware attacks in different sectors in 2021 (*Till November 2021)}
	\label{fig:p1}
	\vspace{-10pt}
\end{figure}

Due to the increasing number of ransomware variants and ransomware attacks, researchers have been earnestly involving themselves to look for efficient ways to improve the scenarios. While some researchers are analyzing the distinctive behaviors of ransomware by executing it in a secure environment called Dynamic Analysis [6]-[11], some researchers are analyzing the ransomware without any execution, referred to as Static Analysis [12], [13]. However, a good number of researchers are combining these two approaches and adopting a Hybrid Analysis Approach [14], [15]. In this research, we have opted for the dynamic analysis approach for its ability to detect and classify ransomware based on behavioral patterns regardless of the code obfuscation techniques deployed by the ransomware programmers [16], [17]. The main contributions of this paper are:

\begin{itemize}
    \item Develop a Web-Crawler, ‘GetRansomware’ to automate collecting the Windows Portable Executable (PE) files of 15 different ransomware families from the ransomware repository. The Web-Crawler is essential to automate searching and downloading the samples and to cut down the manual workload, but no prior works targeted this scenario.
    \item Develop our dataset and conduct feature selection through a two-phase feature engineering process that includes- ‘Feature Extraction’ from the sample binaries, and ‘Feature Selection’ to select the most important features for each ML classifier.
    \item Develop, evaluate and compare the performance of six State-of-the-art Supervised Machine Learning models. Our approach includes utilizing Recursive Feature Elimination with Cross-Validation (RFECV) for selecting the significant features and RandomSearchCV for selecting the optimum hyperparameter values for each ML classifier. Thereby we attempt to optimize each model's performance before the comparison is made.
    \item Present the post-hoc analysis of the best-performing model using ‘SHapley Additive exPlanations’ or SHAP values to ascertain the transparency and trustworthiness of the model’s prediction. This insight presents a better idea about which features are more dominant in detecting and classifying the ransomware families. While explainability has been widely presented in malware detection scenarios, to the best of the authors’ knowledge, till today, no prior works presented their models' explainability that considered only the ransomware families. 
\end{itemize}

The rest of this paper is structured as follows: Section \ref{sec2} discusses the related works. Section \ref{sec3} presents our proposed method. The experimental results and discussion are made in Section \ref{sec4}. Section \ref{sec5} presents our model's explainability. Section \ref{sec6} concludes the paper with the direction for future works.

\section{Related Works} \label{sec2}
Most researchers prefer the dynamic analysis approach because it can delineate the behaviors of the ransomware in a more explicit manner. Maniath et al. [6] analyzed the API call sequence of 157 ransomware and presented an LSTM-based ransomware detection method. Despite securing 96.67\% accuracy, this work lacks complete information about the ransomware families/variants, and the number of benign software used for the experiment. VinayaKumar et al. [7] proposed an MLP-based ransomware detection method focusing on API call frequency and secured 100\%, and 98\% accuracy for binary and multi-class classification respectively. However, they deployed a simple MLP network that failed to distinguish CryptoWall and Cryptolocker ransomware.  Z. Chen et al. [8] used the API Call Flow Graph (CFG) generated from the extracted API sequence of 83 ransomware and 83 benign software. Regardless of securing 98.2\% exactness using the Logistic Regression model, the work is based on a smaller dataset that includes only four ransomware families. Also, graph-similarity analysis requires higher computational power that some systems may fail to provide. Takeuchi et al. [9] used API call sequences extracted from 276 ransomware, and 312 benign files to identify zero-day ransomware attacks. Although the work secured 97.48\% accuracy by deploying the Support Vector Machine, the accuracy of this work decreases while using standardized vector representation because of the less diverse dataset. Using the Intel Pin Tool, Bae et al. [10] extracted the API call sequences from 1000 ransomware, 900 malware, and 300 benign files. Their sequential process includes generating an n-gram sequence, input vector, and Class Frequency Non-Class Frequency (CF-NCF) for every sample before fitting their model. Regardless of obtaining 98.65\% accuracy using the Random Forest classifier, the model’s performance can be improved with the help of deception-based techniques. Hwang et al. [11] analyzed the API call sequence of 2507 ransomware and 3886 benign files. They used two Markov chains, one for ransomware and another for benign software to capture the API call sequence patterns. By using Random Forest, they compensate Markov Chains and control FPR and FNR to achieve better performance. Despite securing 97.3\% accuracy, their model produces high FPR that can be improved with the help of signature-based techniques.

A good number of researchers chose the static analysis approach to detect ransomware. Baldwin and Dehghantanha [12] analyzed the opcode characteristics of 5 crypto-ransomware families and 350 benign samples. Their experiment involved the WEKA AI toolset, and the experimental results showed an accuracy of 96.5\% while recognizing five crypto-ransomware families and benign software by using the Support Vector Machine classifier. However, their work could be improved by extending the dataset and extracting those groups of opcodes identified during the evaluation of attribute selection. Zhang et al. [13] analyzed the opcode-based characteristics of 1787 ransomware of 8 different ransomware families and 100 benign software. Their technique included moving opcode groupings to the N-gram sequence and afterward Term Frequency Inverse Document Frequency (TF-IDF). Five ML classifiers were used with 10-fold cross-validation among which the Random Forest classifier achieved the highest 91.43\% exactness. However, their model could not distinguish Reveton, CryptoWall, and Locky.

Some researchers adopted a hybrid analysis approach that combines the features extracted from the dynamic and static analyses. Subedi et al. [14] used both dynamic and static analysis on the library, assembly, and function calls. Moreover, they came up with a new analysis tool, namely, CRSTATIC which was deployed to build signatures that could classify ransomware families with the help of reverse engineering. However, they analyzed only 450 samples of ransomware for their experiment and their experimental results showed some important correlations among various code components instead of detecting ransomware. Ferrante et al. [15] proposed a crossover way to deal with android ransomware recognition. While the dynamic detection strategy considered memory utilization, system call insights, CPU, and network utilization, the static identification technique utilized the recurrence of opcodes. They analyzed 2386 benign android applications and 672 applications containing ransomware for their experiment and used 3 ML classifiers- J48, Naïve Bayes and Logistic Regression. Although their experimental result showed 100\% accuracy, their test set for the hybrid model included only 9 ransomware applications that were not detected in the static analysis phase but were detected in the dynamic analysis.

Regardless of the relentless effort of the prior research works, we have found some common gaps and challenges, such as a) manual data collection process, b) not having a large or diverse dataset that focuses on both Crypto and Locker types of ransomware, c) using the traditional sandbox environment that often falls short providing in-depth analysis report (i.e., Cuckoo Sandbox), d) modeling binary classification models to detect ransomware from malware or benign software, and e) not presenting the models’ explainability. Therefore, the goal of our research work lies in a) automating the data collection process, b) constructing a balanced dataset consisting of the most potential features extracted through a methodical dynamic analysis of both Crypto and Locker types of ransomware, c) utilizing an efficient sandbox environment that conducts large scale integration with other antivirus engines and provides in-depth analysis reports (i.e., Falcon Sandbox), d) developing efficient and faster Supervised Machine Learning model that can attain a higher degree of accuracy in case of multi-class classification, and e) presenting the explainability of our model to affirm the trustworthiness of the model’s prediction.

\section{Proposed Method} \label{sec3}
Fig. 2 illustrates the process overview of our proposed method which consists of three steps: Data Collection, Feature Engineering, and Classification.

\begin{figure*}[htb]
	\makebox[\linewidth][c]{\includegraphics[angle = 0, clip, trim=0cm 0cm 0cm 0cm, width=0.80\textwidth]{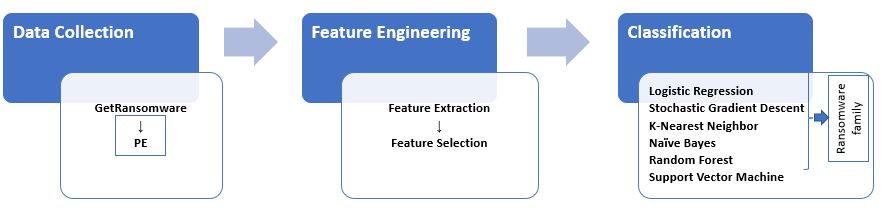}}
	\caption{Process overview of our proposed method}
	\label{fig:p2}
	\vspace{-10pt}
\end{figure*}

\subsection{Data Collection}
We have developed a Web-Crawler, namely, ‘GetRansomware’ to automate searching and downloading Windows Portable Executable (PE) files of ransomware from VirusShare [18]. We have also shared our Web-Crawler on our GitHub repository for public access [19]. About 95\% of our sample ransomware was collected from VirusShare using GetRansomware. The rest of the ransomware samples were collected from theZoo [20] and Hybrid-Analysis.com [21]. Our samples included both Crypto and Locker types of ransomware and every sample was downloaded as a password-protected compressed file from the repositories. In addition, we have collected benign software from the PORTABLE FREEWARE collection [28] which includes portable executables of benign media players, file explorers, file compression tools, file managers, image editors, etc. Table \ref{tab1} presents the description of our dataset.

\begin{table}[htbp]
\caption{Description of Dataset}
\begin{center}
\begin{tabular}{|c|c|}
\hline
\textbf{Ransomware}&\textbf{Number of samples} \\
\hline
Cerber(c1)& 95 \\
\hline
CryptoLocker (c2)& 93 \\
\hline
CryptoWall (c3)& 97\\
\hline
Eris (c4)&	98 \\
\hline
Hive (c5)&	100 \\
\hline
Jigsaw (c6)& 93\\
\hline
Locky (c7)&	95 \\
\hline
Maze (c8)&	100 \\
\hline
Mole (c9)&	100 \\
\hline
Sage (c10)&	100 \\
\hline
Satan (c11)& 100 \\
\hline
Shade (c12)& 97 \\
\hline
TeslaCrypt (c13)&	96 \\
\hline
Virlock (c14)&	93 \\
\hline
WannaCry (c15)&	93 \\
\hline
Benign (c0)& 100 \\
\hline
Total&	1550 \\
\hline
\end{tabular}
\label{tab1}
\end{center}
\end{table}

\subsection{Feature Engineering}
The scarcity of the ransomware dataset is one of the major challenges that hinder the research work in this area [22]. Therefore, we aim to construct our dataset through a thorough feature engineering process to overcome this challenge. The feature engineering step is composed of two phases. The phases are:
\begin{enumerate}
    \item Phase 1: Feature Extraction
    \item Phase 2: Feature Selection
\end{enumerate}

\subsubsection{Phase 1: Feature Extraction}
From the wide range of distinct behavioral features, we have considered utilizing API call frequencies for our classifiers. Application Programming Interface (API) calls are made by the application or program running at a user level to request system services or gain access to the system resources, such as network connection, files, registry, processes, etc. The OS performs the requested services by issuing these calls, and the outcomes are returned to the caller user applications. Thus, API calls made by the ransomware program allow the attackers to obtain the control of the system and perform malicious activities. Since API calls provide a complete picture of a particular program, analyzing API call behavior leads the researchers to better understand the program’s behavior [23], [24]. Therefore, we have considered extracting the API call frequencies by executing the Windows PE files of the ransomware and benign software in a secure virtual environment and utilizing them as distinguishing features for our classification models.

We have analyzed the PE files with the help of Hybrid-Analysis.com [21], powered by the CrowdStrike Falcon Sandbox [25]. To automate submitting malicious binaries, pull the analysis report after the analysis, and perform advanced or required search queries on the database, Falcon Sandbox provides a free, convenient, and efficient API key that one can obtain from an authorized user account. For analysis, we have used our API key and Falcon Sandbox Python API Connector- VxAPI wrapper [26] to automatically submit the binaries from our system. After submission, Falcon Sandbox runs the binaries in a Virtual Machine (VM) and captures the run-time behaviors as illustrated in Fig. 3. Later, it shows the analysis results on the web interface.

Contrary to the prior research works where the analysis tasks were done using the Cuckoo Sandbox [6], [7], [9], [11], we have analyzed the PE files using the Falcon Sandbox that used a VM (Windows 7 64-bit) to execute the PE files. Falcon Sandbox can analyze a wide variety of files with or without dependencies, such as PE (i.e., .com, .exe, .dll, etc.), Office (i.e., .doc, .docx, .ppt, .pptx, etc.), and PDF, etc. Also, Falcon Sandbox incorporates many other services, such as VirusTotal, Thug honeyclient, OPSWAT Metadefender, TOR, NSRL (Whitelist), Phantom, and a large number of antivirus engines to provide an integrated and accurate analysis report compared to other sandboxes. While executing the binaries, we have set run-time to the maximum available duration in the Falcon Sandbox to deal with the delayed execution techniques deployed by the attackers. The total time for the analysis was (1550 PE files * 7 minutes) = 181 hours = 8 days approximately. Next, we obtained the analysis report by using our API key. As we aim to focus on the API call behaviors, we have only sorted and computed the frequency of each API call. At the end of the Feature Extraction phase, we obtained our dataset 'RansomClass' consisting of the frequencies of 68 distinct API calls commonly called by the 15 different ransomware families and benign software as presented in Table \ref{tab2}. 

\begin{figure}[htb]
	\makebox[\linewidth][c]{\includegraphics[angle = 0, clip, trim=0cm 0cm 0cm 0cm, width=0.5\textwidth]{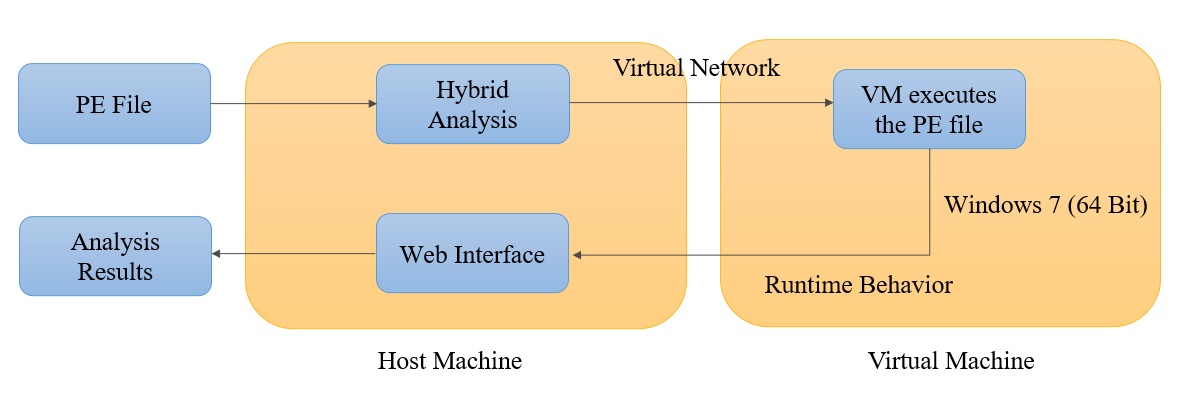}}
	\caption{Block diagram of the PE file execution process}
	\label{fig:p3}
	\vspace{-10pt}
\end{figure}

\begin{table}[htbp]
\caption{List of API Calls}
\begin{tabular}{|c c|}
\hline
\multicolumn{2}{|c|}{\textbf{Name of the API Calls}}\\
\hline
1.	FindWindowExW & 2.	LdrGetDllHandle \\
3.	NtAdjustPrivilegesToken & 4.	NtAlertThread \\
5.	NtAllocateVirtualMemory & 6.	NtAlpcSendWaitReceivePort \\
7.	NtConnectPort & 8.	NtCreateEvent \\
9.	NtCreateFile & 10.	NtCreateKey \\
11.	NtCreateKeyEx & 12.	NtCreateMutant \\
13.	NtCreateSection & 14.	NtCreateThreadEx \\
15.	NtCreateUserProcess & 16.	NtDelayExecution \\
17.	NtDeleteValueKey & 18.	NtDeviceIoControlFile \\
19.	NtEnumerateKey & 20.	NtEnumerateValueKey \\
21.	NtFsControlFile & 22.	NtGetContextThread \\
23.	NtMapViewOfSection & 24.	NtNotifyChangeKey \\
25.	NtOpenDirectoryObject & 26.	NtOpenEvent \\
27.	NtOpenFile & 28.	NtOpenKey \\
29.	NtOpenKeyEx & 30.	NtOpenMutant \\
31.	NtOpenProcess & 32.	NtOpenProcessToken \\
33.	NtOpenSection & 34.	NtOpenThreadToken \\
35.	NtProtectVirtualMemory & 36.	NtQueryAttributesFile \\
37.	NtQueryDefaultLocale & 38.	NtQueryDirectoryFile \\
39.	NtQueryInformationFile & 40.	NtQueryInformationProcess \\
41.	NtQueryInformationToken & 42.	NtQueryKey \\
43.	NtQueryObject & 44.	NtQuerySystemInformation \\
45.	NtQueryValueKey & 46.	NtQueryVirtualMemory \\
47.	NtQueryVolumeInformationFile & 48.	NtReadFile \\
49.	NtReadVirtualMemory & 50.	NtRequestWaitReplyPort \\
51.	NtResumeThread & 52.	NtSetContextThread \\
53.	NtSetInformationFile & 54.	NtSetInformationKey \\
55.	NtSetInformationProcess & 56.	NtSetInformationThread \\
57.	NtSetSecurityObject & 58.	NtSetValueKey \\
59.	NtTerminateProcess & 60.	NtTerminateThread \\
61.	NtUnmapViewOfSection & 62.	NtWaitForMultipleObjects \\
63.	NtWriteFile & 64.	NtWriteVirtualMemory \\
65.	NtYieldExecution & 66.	OpenSCManager \\
67.	OpenServiceW & 68.	SetWindowsHookEx \\
\hline
\end{tabular}
\label{tab2}
\end{table}

\subsubsection{Phase 2: Feature Selection}

 Feature selection is one of the potential steps toward building an efficient ML model because it reduces the probability of overfitting by removing redundant data, improves the model’s accuracy by eliminating irrelevant features, and thereby reduces training time. At the beginning of the Feature Selection phase, we have evenly divided (stratified train-test split) our dataset into train data (75\%) and test data (25\%) to avoid data leakage. Next, we have applied Recursive Feature Elimination with Cross-Validation (RFECV) to our train data for selecting the most significant features for each ML classifier. RFECV is a wrapper-style feature selection method that wraps a given ML model and selects the optimal number of features for that model by recursively eliminating $0-n$ features in each loop. Next, it selects the best performing subset of features based on the accuracy or the score of cross-validation. Besides, RFECV also removes the dependencies and collinearities existing in the model. By using RFECV, we have selected 6 distinct subsets of features for 6 ML classifiers. Table \ref{tab3} presents the list of features that have been selected by RFECV for each ML classifier. To represent the names of the selected features, the serial numbers from Table \ref{tab2} are used in Table \ref{tab3}. These features have been selected by setting ‘min\_features\_to\_select’ as 34 (half of the features), cv=5, and ‘scoring’= ‘accuracy’ so that RFECV would select at least half of the features based on the optimum accuracy over the 5-fold cross-validation. 

\begin{table}[htbp]
\caption{List of features selected by RFECV for each ML classifier}
\begin{center}
\begin{tabular}{|c|c|c|}
\hline
Classifier&	Selected Features&	Total \\
\hline
LR&	\makecell{1,2,4-8,10,11,14-16,18-22,24,28,29,31-35,\\ 37-39,42,43,45,46,48,53,54,56,59,61,63,64,67}& 41 \\
\hline
SGD& \makecell{1,2,4-7,10,11,14-16,19,20,23-39,43, \\45-48,51,53,54,56,59,61,63,64,67}& 44 \\
\hline
KNN&	\makecell{4-8,12-16,19,23-25,27-36,39-41, \\43-48,50,51,53,54,56,59,61,63,64} & 42 \\
\hline
NB&	\makecell{1-4,6,10-12,14,15,17,19-22,24,28,30-34,36-38, \\ 41-44,48,49,51,53,54,56,58,59,61-65,67,68} & 44 \\
\hline
RF&	\makecell{2,3,5,6,8-11,13,14,16,18-20,24,27-36,38-48, \\ 50,51,54-56,59,62,63}& 44 \\
\hline
SVM& 	\makecell{1,2,4-7,9,10,13-16,19-20,23-25,27-40,43, \\45-48,50,51,53,54,56,59,61,63,64} & 44 \\
\hline
\end{tabular}
\label{tab3}
\end{center}
\end{table}

\subsection{Classification}
We have employed Supervised Machine Learning algorithms to detect and classify 15 ransomware families into corresponding categories. Supervised Machine Learning algorithms are trained on the labeled dataset to make a decision in response to the unseen test dataset. As our work focuses on detecting and classifying 15 ransomware families, we have employed the following 6 state-of-the-art Supervised Machine Learning algorithms that are widely used for both binary and multi-class classification problems as per requirement:

Logistic Regression (LR): a type of statistical analysis that predicts the probability of a dependent variable from a set of independent variables using their linear combination.

Stochastic Gradient Descent (SGD): is an optimization algorithm to find the model parameters by updating them for each training data so that the best fit is reached between predicted and actual outputs.

K-Nearest Neighbor (KNN): estimates the likelihood of a new data point being a member of a specific group by measuring the distance between neighboring data points and the new data point. 

Naïve Bayes (NB): is based on Bayes’ theorem and predicts the probability of an instance belonging to a particular class.

Random Forest (RF): constructs multiple decision trees and finally determines the class selected by the maximum number of trees. 

Support Vector Machine (SVM): takes one or more data points from different classes as inputs and generates hyperplanes as outputs that best distinguish the classes.

Since we aim to do multi-class classification and some classifiers are only designed for binary classification problems (i.e., Logistic Regression, Support Vector Machine, etc.), we cannot directly apply these classifiers for multi-class classification problems. Therefore, heuristic methods can be applied to divide a multi-class classification problem into several binary classification problems. There are two types of heuristic methods, such as One-vs-Rest (OvR) which splits the dataset into one class against all other classes each time, and One-vs-One (OvO) which splits the dataset into one class against every other class each time. We have applied the OvR method for our experiment to reduce the time and computational complexities. All these classifiers are built along with ‘RandomSearchCV’- a hyperparameter optimization technique, to find the best combination of hyperparameters for maximizing the performance of the models’ output in a reasonable time. Instead of exhaustively searching for the optimal values of the hyperparameters through a manually determined set of values (i.e., Grid Search), RandomSearchCV randomly searches the grid space and selects the best combination of hyperparameter values based on the accuracy or the score of cross-validation. Since we have used RFECV for feature selection and RandomSearchCV for hyperparameter optimization, the nested cross-validation technique has been applied to build our model.

\section{Experimental Results} \label{sec4}

We have evaluated our models in terms of Precision, Recall, F1-score, Accuracy, and Area Under the Receiver Operator Characteristic (AUROC). These performance metrics are measured as follows:

\begin{equation}
Precision = \frac{TP}{TP+FP} \label{eq:precision}
\end{equation}

\begin{equation}
Recall = \frac{TP}{TP+FN} \label{eq:recall}
\end{equation}

\begin{equation}
F1{\text -}score = \frac{2 \times Precision \times Recall}{Precision + Recall} \label{eq:f1}
\end{equation}

\begin{equation}
Accuracy = \frac{TP + TN}{TP + TN + FP + FN} \times 100 \label{eq:acc}
\end{equation}
\noindent where, TP = True Positives, FP = False Positives, TN = True Negative, FN = False Negatives.

Table \ref{tab4} describes the experimental results showing that LR outperforms other classifiers securing 99.15\% overall accuracy. Although KNN achieves 85.65\% accuracy, the processing time of this classifier is less than other classifiers as presented in Table \ref{tab5}. To examine the degree of separability of our best-performed machine learning model, we have constructed Area Under the Receiver Operator Characteristic (AUROC) curve of the LR classifier as shown in Fig. 4. We construct the AUROC curve by plotting the True Positive Rate or Sensitivity against the False Positive Rate or (1-Specificity). For threshold = 0.5, the average AUROC of LR is 99.55. This is an ideal situation for LR which indicates that the model has a 99.55\% probability of predicting the correct class against the rest. Therefore, we can ascertain that LR is our best-performing classifier.

\begin{table}[htbp]
\caption{Performance ($P_{avg}$) comparison of LR, SGD, KNN, NB, RF, and SVM Classifiers}
\begin{center}
\begin{tabular}{|c|c|c|c|c|c|c|}
\hline
$P_{avg}$&	LR& 	SGD&	KNN&	NB&	RF&	SVM \\
\hline
Accuracy&	99.15&	85.26&	85.65&	94.96&	82.38&	94.23 \\
\hline
Precision&	99.22&	95.94&	93.47&	98.04&	99.86&	99.13 \\
\hline
Recall&	99.15&	87.62&	85.65&	94.96&	82.38&	94.23 \\
\hline
F1-score&	99.15&	91.14&	88.04&	96.28&	89.31&	96.37 \\
\hline
AUROC&	99.55&	93.67&	92.55&	97.41&	91.19&	97.09 \\
\hline
\end{tabular}
\label{tab4}
\end{center}
\end{table}

\begin{table}[htbp]
\caption{Classifier’s processing time comparison}
\begin{center}
\begin{tabular}{|c|c|}
\hline
Classifier&	Processing Time (in seconds) \\
\hline
LR&	52.44 \\
\hline
SGD&	53.65 \\
\hline
KNN&	51.34 \\
\hline
NB&	53.77 \\
\hline
RF&	54.19 \\
\hline
\end{tabular}
\label{tab5}
\end{center}
\end{table}

\begin{figure}[htb]
	\makebox[\linewidth][c]{\includegraphics[angle = 0, clip, trim=0cm 0cm 0cm 0cm, width=0.45\textwidth]{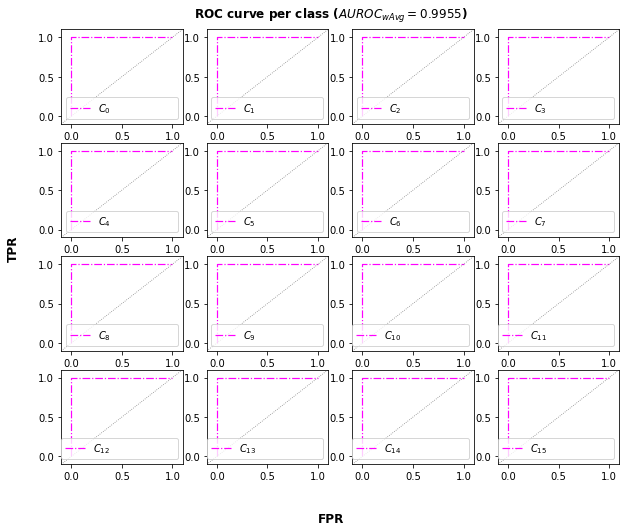}}
	\caption{AUROC curve of Logistic Regression}
	\label{fig:p4}
	\vspace{-10pt}
\end{figure}

\section{Model Explainability} \label{sec5}
Although some predictive models may not require explanation because of their usage in a low-risk real-world environment, well-studied problem statements, and the availability of sufficient practical solutions, some models that deal with the real-world high-risk environment (i.e., ransomware detection/classification) need explanation. When it comes to the performance evaluation of any model, knowing both ‘What’ and ‘Why’ the models have taken these decisions is equally important. These insights lead us to dive deeper into the problems, explore the most important features that affect the model’s prediction both locally and globally and thereby build the trustworthiness of the model’s predictions. We present the explainability of our model’s prediction with the help of ‘SHapley Additive exPlanations’ or SHAP values [27]. SHAP is based on the coalition game theory that measures each feature’s individual contribution to the final output while conserving the sum of contributions being the same as the final result. Contrary to the other explanation techniques, such as permutation feature importance, LIME, etc., SHAP provides both global and local explainability. Besides, we can use SHAP values to explain a wide variety of models, such as DeepExplainer to explain Deep Neural Networks (i.e., Multi-Layer Perceptron, Convolutional Neural Network, etc.), TreeExplainer to explain tree-based models (i.e., Random Forest, XGBoost, etc.), and KernelExplainer to explain any model, etc.

To illustrate both local and global explainability, we have used KernerExplainer to explain our Logistic Regression model.

\subsection{Local Interpretation}
For our classification model, the SHAP value is regarded as a 2-D array where the columns represent the features used in the model and the rows represent individual predictions predicted by the model. So, each SHAP value in this array indicates a specific feature’s contribution to the output of that row’s prediction as shown in Fig. 5. Here, a positive SHAP value specifies that a feature is positively pushing the base value or expected value to the model output. On the other hand, the negative SHAP value specifies that the feature is negatively pushing the base value to the model output. The base value or the mean model output is computed based on the train data. By using the Force plot, we can visualize this explanation for an individual prediction as shown in Fig. 6. For one instance, the features with higher SHAP values (red) are positively pushing the base value to the model output, and the features with lower SHAP values (blue) are negatively pushing the base value to the model output.

\begin{figure}[htb]
	\makebox[\linewidth][c]{\includegraphics[angle = 0, clip, trim=0cm 0cm 0cm 0cm, width=0.45\textwidth]{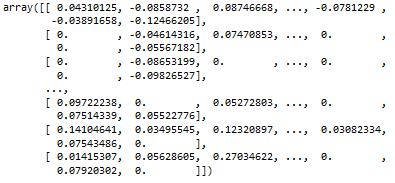}}
	\caption{Array of SHAP values}
	\label{fig:p5}
	\vspace{-10pt}
\end{figure}

\begin{figure*}[htb]
	\makebox[\linewidth][c]{\includegraphics[angle = 0, clip, trim=0cm 0cm 0cm 0cm, width=0.80\textwidth]{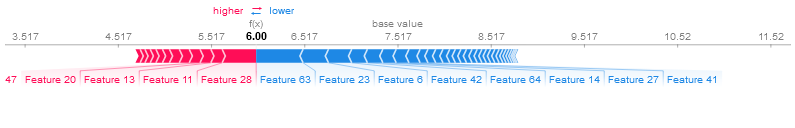}}
	\caption{Force plot for single instance}
	\label{fig:p6}
	\vspace{-10pt}
\end{figure*}

\subsection{Global Interpretation}
Passing the array of SHAP values to a ‘summary\_plot’ function creates a feature importance plot as shown in Fig. 7. It illustrates 20 highly contributing features to the model output. Here, the x-axis denotes the mean of the absolute SHAP value for each feature which indicates the total contribution of the feature to the model and the y-axis denotes the features used for the classification. The features are organized in descending order from top to bottom by how strongly they influence the model’s decision. For our LR model, the frequency of NtAlpcSendWaitReceivePort is the most important feature with a higher mean absolute SHAP value followed by NtFsControlFile, NtWriteVirtualMemory, etc.

\begin{figure}[htb]
	\makebox[\linewidth][c]{\includegraphics[angle = 0, clip, trim=0cm 0cm 0cm 0cm, width=0.48\textwidth]{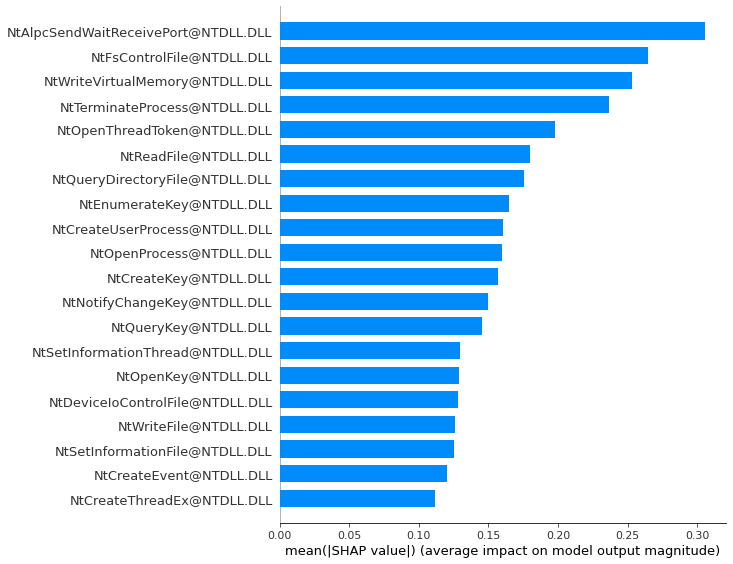}}
	\caption{Summary plot showing feature importance}
	\label{fig:p7}
	\vspace{-10pt}
\end{figure}

In addition to the summary plot, we have created a ‘beeswarm’ plot as presented in Fig. 8 which shows the distribution of the features' impact on the model output. The x-axis denotes the SHAP value and the y-axis denotes each feature used for the classification. Every point on this plot stands for one SHAP value for one feature and prediction. For our LR model, the lower values (blue) of NtAlpcSendWaitReceivePort have a negative impact on the model output, while the lower values of NtFsControlFile have a positive impact on the model output.

\begin{figure}[htb]
	\makebox[\linewidth][c]{\includegraphics[angle = 0, clip, trim=0cm 0cm 0cm 0cm, width=0.48\textwidth]{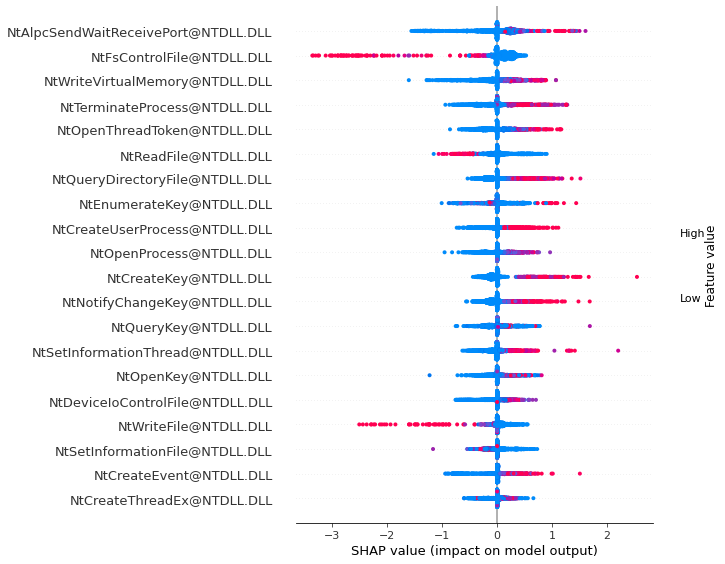}}
	\caption{Beeswarm plot showing directionality impact}
	\label{fig:p8}
	\vspace{-10pt}
\end{figure}

After obtaining these insights from the explanation of our LR model, we conducted another distinct experiment to reevaluate the performance of our LR model including only 20 highly contributing features as shown in Fig. 7 and Fig. 8. As illustrated in Table \ref{tab6}, our model could distinguish the ransomware families with 95.48\% overall accuracy based on those features. This infers that the rest 21 features selected by RFECV for the LR classifier are also contributing to the model achieving its best 99.15\% overall accuracy.

\begin{table}[htbp]
\caption{Performance evaluation of LR model including 20 highly contributing features}
\begin{center}
\begin{tabular}{|c|c|}
\hline
Performance&	LR (including only the top 20 features) \\
\hline
Accuracy&	95.48 \\
\hline
Precision&	96.06 \\
\hline
Recall&	95.48 \\
\hline
F1-score&	95.45 \\
\hline
AUROC&	97.59 \\
\hline
\end{tabular}
\label{tab6}
\end{center}
\vspace{-10pt}
\end{table}

To evaluate our model’s performance concerning the existing classification methods, we have compared the accuracy of the previous API call-based ransomware detection and classification approaches with the proposed method. Both API call frequency and sequence-based detection methods were surveyed from the literature. Table \ref{tab7} illustrates the comparison of the proposed model with those previous works. It is evident that the proposed method provides higher accuracy while having the advantage of multi-class classification with model explainability. In addition, we have developed a Web-Crawler to automate collecting 15 different ransomware families unlike [7] where only 7 ransomware families were utilized for the experiment. Besides, we have conducted feature engineering and utilized hyperparameter optimization techniques to optimize each model's performance contrary to [6]-[11]. Therefore, our approach is unique and efficient in multi-class classification.

\begin{table}[htb]
\caption{Performance comparison with existing ransomware classification models}
\begin{center}
\begin{tabular}{|p{45pt}|p{25pt}|p{50pt}|p{30pt}|p{45pt}|}
\hline
Reference&	Feature&	Classifier&	Multiclass&	Accuracy (\%) \\
\hline
Maniath et al. [6]&	API call&	Long Short-Term Memory&	No&	96.67 \\
\hline
VinayaKumar et al. [7]&	API call&	Multilayer Perceptron&	Yes&	98.00 \\
\hline
Z.-G. Chen et al. [8]&	API call&	Logistic Regression&	No&	98.20 \\
\hline
Takeuchi et al. [9]&	API call&	Support Vector Machine&	No&	97.48 \\
\hline
Bae et al. [10]&	API call&	Random Forest&	Yes&	98.65 \\
\hline
Hwang et al. [11]&	API call&	Markov Chain, Random Forest&	No&	97.30 \\
\hline
Proposed Method&	API call& 	Logistic Regression&	Yes&	99.15 \\
\hline
\end{tabular}
\label{tab7}
\end{center}
\vspace{-10pt}
\end{table}

\section{Conclusion} \label{sec6}
Ransomware has become the most popular weapon to cybercriminals recently for its easy way to earn money through untraceable cryptocurrency. In this research work, we have developed a Web-Crawler to automate searching and downloading both Crypto and Locker types of ransomware. By executing the binaries of ransomware and benign software using an advanced sandbox tool, we extracted API call frequencies, constructed our dataset, and selected features through a two-phase feature engineering process. Next, we have applied six state-of-the-art Supervised Machine Learning models to detect and classify ransomware families. Our approach includes utilizing Recursive Feature Elimination with Cross-Validation (RFECV) for selecting the significant features and RandomSearchCV for selecting the optimum hyperparameter values for each ML classifier. Thereby we attempt to optimize each model's performance. The results reveal that Logistic Regression can efficiently classify ransomware into their corresponding families securing 99.15\% overall accuracy. Finally, instead of relying on the ‘Black box’ characteristic of the Machine Learning models, we also present the explainability of our best-performing model using SHAP values to ascertain the transparency and trustworthiness of the model’s prediction. In the future, we aim to enlarge our dataset by including more samples and potential distinguishing features. In addition, we plan to apply state-of-the-art deep neural networks with explainability for ransomware classification.

\vspace{12pt}
\color{red}

\end{document}